\documentclass[aps,prb,twocolumn,superscriptaddress]{revtex4-1}
\usepackage{amsmath}
\usepackage{amssymb}
\usepackage{graphicx}
\usepackage{multirow}

\begin{document}

\title{Deviations from Matthiessen rule and resistivity saturation effects in Gd and Fe}

\author{J. K. Glasbrenner}
\altaffiliation[Present address: ]{Code 6393, National Research Council/Naval Research Laboratory, Washington, DC 20375, USA}
\affiliation{Department of Physics and Astronomy and Nebraska Center for Materials and Nanoscience, University of Nebraska--Lincoln, Lincoln, Nebraska 68588, USA}
\author{B. S. Pujari}
\altaffiliation[Present address: ]{Centre for Modeling and Simulation, University of Pune, Ganeshkhind, Pune 411007, India}
\affiliation{Department of Physics and Astronomy and Nebraska Center for Materials and Nanoscience, University of Nebraska--Lincoln, Lincoln, Nebraska 68588, USA}
\author{K. D. Belashchenko}
\affiliation{Department of Physics and Astronomy and Nebraska Center for Materials and Nanoscience, University of Nebraska--Lincoln, Lincoln, Nebraska 68588, USA}
\affiliation{Kavli Institute for Theoretical Physics, University of California, Santa Barbara, California 93106, USA}

\date{\today}

\begin{abstract}
According to earlier first-principles calculations, the spin-disorder contribution to the resistivity of rare-earth metals in the paramagnetic state is strongly underestimated if Matthiessen's rule is assumed to hold.
To understand this discrepancy, the resistivity of paramagnetic Fe and Gd is evaluated by taking into account both spin and phonon disorder. Calculations are performed using the supercell approach within the linear muffin-tin orbital method. Phonon disorder is modeled by introducing random displacements of the atomic nuclei, and the results are compared with the case of fictitious Anderson disorder. In both cases the resistivity shows a nonlinear dependence on the square of the disorder potential, which is interpreted as a resistivity saturation effect. This effect is much stronger in Gd than in Fe. The non-linearity makes the phonon and spin-disorder contributions to the resistivity non-additive, and the standard procedure of extracting the spin-disorder resistivity by extrapolation from high temperatures becomes ambiguous. An ``apparent'' spin-disorder resistivity obtained through such extrapolation is in much better agreement with experiment compared to the results obtained by considering only spin disorder. By analyzing the spectral function of the paramagnetic Gd in the presence of Anderson disorder, the resistivity saturation is explained by the collapse of a large area of the Fermi surface due to the disorder-induced mixing between the electronic and hole sheets.
\end{abstract}

\maketitle

\section{Introduction}
\label{section-intro}

The electric resistivity of magnetic metals is due to several scattering mechanisms, including scattering on impurities, lattice vibrations, and spin fluctuations.\cite{coles,mott,vonsovskiibook} While the impurity and phonon scattering are well understood both on the general level \cite{zimanbook} and quantitatively, \cite{impurity1,impurity2,impurity3,Maksimov} spin-disorder scattering has not been studied based on the first-principles electronic structure theory until recently. \cite{alexsdr,jamessdr} Understanding of this scattering mechanism is important, because it provides quantitative information about the character of thermal spin fluctuations in metals.\cite{moriyabook}

The interpretation of resistivity measurements in magnetic metals usually assumes that Matthiessen's rule holds. \cite{weissmarotta}
Under this assumption it makes sense to talk about the individual spin-disorder contribution to the resistivity, which does not depend on the intensity of other scattering mechanisms. If the local moments are temperature-independent, this contribution saturates in the paramagnetic state, which allows one to fit and subtract out the residual and phonon contributions. The remaining part obtained in this way will be referred to below as the apparent spin-disorder resitivity (SDR).

To calculate the SDR from first-principles, the most general approach is to construct supercells representing an ensemble of spin disorder configurations, average the Landauer-B\"{u}ttiker conductance over this ensemble, and extract the resistivity from the scaling of the result with the dimensions of the supercell. This approach has been applied to transition metals Fe
and Ni\cite{alexsdr} and to the Gd-Tm series of heavy rare-earth metals.\cite{jamessdr}
A simpler procedure is to calculate the resistivity using the Kubo-Greenwood formula applied to the disordered local moment (DLM) state, \cite{gyorffydlm} which represents the coherent potential approximation (CPA) applied to the paramagnetic state. The application of this procedure is similar to the calculation of the residual resistivity of substitutional alloys.\cite{dlmlinresp1,dlmlinresp2}  The results for transition metal ferromagnets \cite{josefdlm} and heavy rare-earth metals \cite{jamessdr} were found to agree very well with the supercell calculations.

For transition metals and alloys, calculated SDR is generally in good agreement with experimental data.\cite{alexsdr,josefdlm}
In contrast, for heavy rare-earth metals in the Gd-Tm series the calculated SDR is systematically underestimated.\cite{jamessdr}
For heavier elements in the series the agreement with experiment is significantly improved by applying the $(S+1)/S$ quantum correction, which corresponds to the limit of weak spin-orbit coupling. The justification for this choice is lacking, absent a consistent description of the conduction electron scattering on localized spins in the regime when hybridization is comparable to spin-orbit multiplet splittings. This uncertainty complicates the comparison of the calculated SDR with experimental data for the heavier elements. However, for lighter elements with large spin moments, particularly Gd, a large underestimation of the resistivity by more than a factor of 2 can not be explained by any kind of quantum correction, and its origin should therefore be sought in the details of the electronic structure and scattering. In particular, the validity of Matthiessen's rule in the presence of strong spin and phonon disorder should be brought into question.

In this paper we extend our supercell approach \cite{alexsdr,jamessdr} to evaluate the resistivity in the presence of both spin and phonon disorder. We apply this method to Fe and Gd and find significant deviations from Matthiessen's rule with increasing disorder, which are particularly strong for Gd and indicate a hidden resistivity saturation effect. As a result, the SDR calculated at zero lattice displacements becomes much smaller than the value obtained by extrapoling the high-temperature data, which provides an explanation for the apparent underestimation of SDR in previous calculations neglecting the phonons. The rest of the paper is organized as follows. In Sec.\ \ref{section-methods} we describe the methods used in the calculations of the resistivity, and in Sec.\ \ref{section-transportresults} the results for Fe and Gd are presented. In Sec.\ \ref{section-electronicstructure} we analyze the electronic structure of Gd in the presence of disorder and identify the origin of the resistivity saturation effect. The conclusions are drawn in Sec.\ \ref{section-conclusions}.

\section{Computational methods}
\label{section-methods}

Atomic displacements can be included explicitly in supercell calculations. Electron scattering on such frozen thermal lattice disorder is a good representation of phonon scattering at temperatures that are not too low compared to the Debye temperature. With an uncorrelated Gaussian distribution for the lattice displacements, this approach was recently employed to study Gilbert damping.\cite{kellyphonons} The lattice displacements can also be determined more realistically from the Born model or \emph{ab-initio} molecular dynamics simulations.\cite{alfephonon} It is also possible to include uncorrelated atomic displacements within the CPA. \cite{Gonis1,Gonis2,Ebert} We have followed the approach of Ref.\ \onlinecite{kellyphonons} in this work.

All calculations were performed using the tight-binding linear muffin-tin orbitals (LMTO) method within the atomic sphere approximation and with the local spin density approximation (LSDA) for the exchange-correlation potential. Spin disorder is introduced by randomly assigning the direction of the local magnetic moment vector on each atom in the supercell. \cite{alexsdr,jamessdr}
The effects of spin and lattice disorder can thus be studied on the same footing.

We have considered both $\alpha$ and $\gamma$ phases of Fe, setting the lattice parameters to their experimental values: 2.8655 \AA\ for $\alpha$ and 3.6394 \AA\ for $\gamma$-Fe, the latter measured close to the $\alpha$-$\gamma$ phase transition. \cite{felatparam}
For hcp Gd we also used the experimental parameters $a=3.629$ \AA\ and $c/a=1.597$.
The conduction electrons were represented by the basis set including $s$, $p$, and $d$ electrons.
For Gd the $4f$ electrons were treated in the ``open core'' approximation as in our earlier calculations. \cite{jamessdr}

\begin{figure}[htb]
\centering \includegraphics[width=0.45\textwidth]{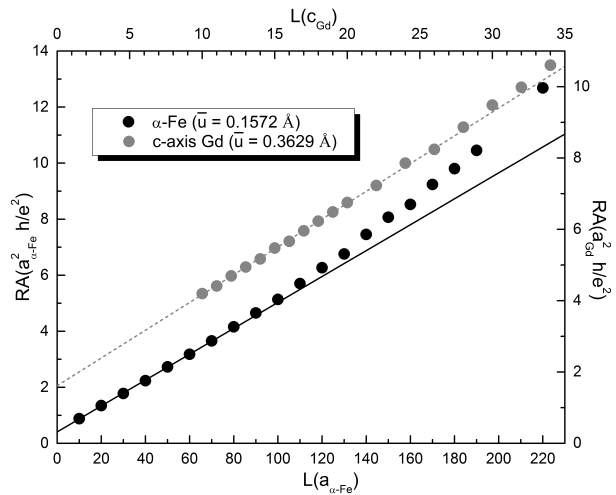}
\caption{The area-resistance product as a function of the active disordered region length for two separate sets of calculations. The black circles (read using bottom and left axes) are calculations with collinear ferromagnetic $\alpha \text{-Fe}$ and a
phonon mean-square displacement $\bar{u} = 0.1572 \text{ \AA}$, and the gray
circles (read using top and right axes) are calculations of Gd with random
noncollinear spin disorder, current flowing parallel to the c-axis, and phonon
mean-square displacement $\bar{u} = 0.3629 \text{ \AA}$}  \label{pic-ravsldata}
\end{figure}

The conductance of each supercell was calculated using the Landauer-B\"{u}ttiker approach. The results for different lengths of the disordered scattering region were fitted to Ohm's law as shown in Fig.\ \ref{pic-ravsldata}, and the resistivity is obtained from the slope of this dependence. For longer lengths of the scattering region the system becomes effectively one-dimensional and Ohmic scaling breaks down due to Anderson localization.\cite{localization} As shown in Fig.\ \ref{pic-ravsldata}, the fits to Ohm's law were based on the range of lengths where the localization effects are negligible.

The resistivity is isotropic for Fe, while for hcp Gd the tensor has two independent
components for current flowing parallel and perpendicular to the $c$ axis.
We used supercells with a cross-section of $4a\times4a$  (16 atoms per monolayer) for $\alpha$-Fe; $3a \times 3a$ (18 atoms per monolayer) for $\gamma$-Fe; $4a \times 4a$ (16 atoms per monolayer) for Gd with current flowing along the $c$ axis; and $3\sqrt{3}a\times2c$ (12 atoms per monolayer) for Gd with current along an in-plane translation vector.
The Brillouin zone integration was performed using meshes that ranged from $15\times15$ for $\alpha$-Fe with vector spin disorder to $25 \times 25$ for $\alpha$-Fe with collinear (Ising) spin disorder. The conductance was
averaged over 15 or 30 disorder configurations when the root mean-squared atomic displacement $\bar u$ was less than or greater than $0.08a$, respectively.

For $\gamma$-Fe and Gd with the in-plane transport direction, the dependence of the resistivity on the magnitude of the local magnetic moments was checked by using the atomic potentials taken from the ferromagnetic or the paramagnetic state as input for the transport calculations. The paramagnetic state was modeled using the DLM approach in this case.\cite{gyorffydlm} For $\alpha$-Fe and Gd with the transport along the $c$ axis we only used the potentials from the ferromagnetic state.

For further analysis, we calculated the $c$ axis resistivity of Gd with artificial Anderson disorder introduced instead of the lattice displacements. This was done by adding random shifts to the atomic potentials of different sites (on-site band-center parameters $C$ and linearization energies $E_\nu$ in LMTO), which were distributed uniformly in the range of $(-\Delta,\Delta)$. We performed two sets of calculations for this system, one with random vector spin disorder and atomic potentials from the ferromagnetic state (averaging over 15 disorder configurations), and another with zero magnetic moments on all sites (30 configurations).

The densities of states (DOS) of Gd with spin and lattice disorder were calculated using a 64-atom supercell (4 hexagonal
monolayers with 16 atoms per monolayer). The atomic potentials were taken from the ferromagnetic state (local moment $m=7.72 \mu_B$). Seven random vector spin disorder configurations were generated for averaging, and random lattice displacements of different amplitudes were introduced as described above. The partial spin-dependent density of states (DOS) was then calculated for each atom in the local reference frame in which the $z$ axis is parallel to the direction of the local magnetic moment. This partial DOS was then averaged over all atoms and disorder configurations.

The Bloch spectral functions of Gd with spin and Anderson disorder were calculated using the standard technique within the CPA. \cite{Turek-book} Here, instead of a uniform distribution of the disorder potential, we assumed that the local potential shift randomly takes two values, $\Delta$ and $-\Delta$. The two spin orientations combined with two values of the potential shift are then formally treated in CPA as a four-component random alloy.

\section{Electrical resistivity of F\lowercase{e} and G\lowercase{d}}
\label{section-transportresults}

\begin{figure}[htb]
\centering \includegraphics[width=0.45\textwidth]{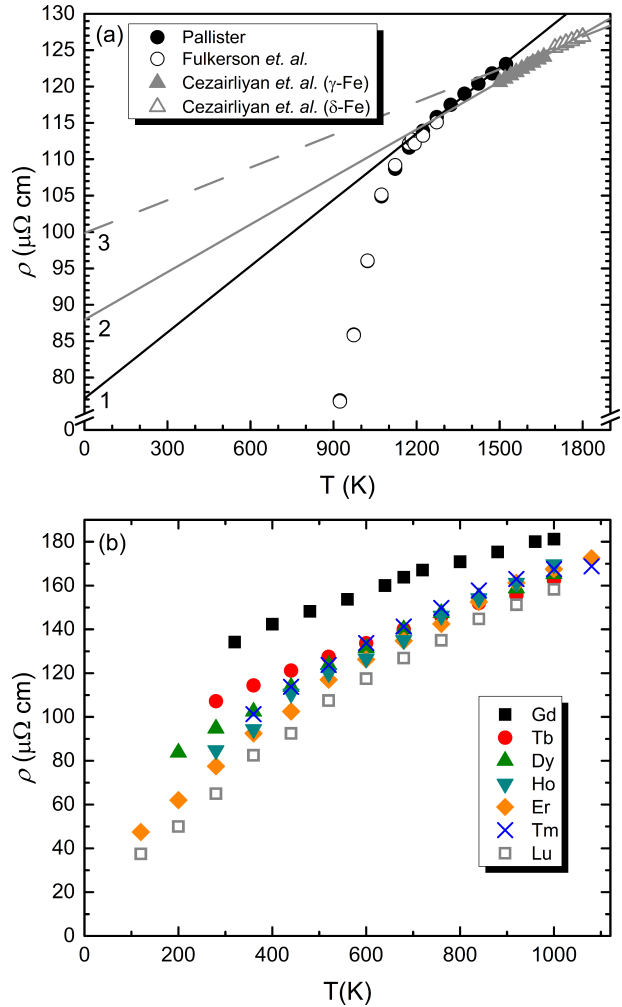} \caption{Electrical
resistivity data taken from experiment. (a) Fe resistivity measurements compiled from Refs.~\onlinecite{pallister,fulkerson,cezairliyan}.
The three lines correspond to fits to data from (1) Pallister;\cite{pallister}
(2) and (3) Cezairliyan \emph{et.~al.}\cite{cezairliyan} The temperature range,
slopes, and intercepts of the fits are in Table~\ref{table-exptfedata}. (b) High-temperature resistivity data for polycrystalline rare earth metals compiled from Ref.\ \onlinecite{vedernikov}.}
\label{pic-exptdata}
\end{figure}

It has become common practice to determine the SDR by extrapolating the high-temperature resistivity data back to zero temperature. This procedure relies on the assumption that spin-disorder and phonon scattering processes are independent, which is a good approximation as long as the electronic states retain their quasiparticle character and their band structure is weakly affected by disorder. If these conditions are satisfied, a linear temperature dependence of resistivity is expected at temperatures above the Debye temperature. Deviations from linearity are, however, rather common.
Consider the resistivity measurements for Fe, \cite{pallister,fulkerson,cezairliyan} which are assembled in Fig.~\ref{pic-exptdata}a. Our fits to these data are included in the figure and summarized in Table~\ref{table-exptfedata}. The $\alpha$ ($T < 1180$  K) and $\delta$ ($T > 1680$ K) phases of Fe are crystallographically identical, and the corresponding resistivity data should lie on the same smooth curve. It seems clear that this curve deviates significantly from the straight line in the paramagnetic region. In particular, the intercept of fit 3 is 1.3 times larger and its slope 2 times smaller compared to fit 1 (see Table~\ref{table-exptfedata}). The same trend (sublinear temperature dependence) has also been observed for polycrystalline samples of heavy rare-earth metals measured between room temperature and 1000 K,\cite{vedernikov} which we have compiled in Fig.\ \ref{pic-exptdata}b. As an example, the slope of the resistivity data for paramagnetic Er decreases by nearly a factor of 2 over this temperature range. A similar deviation from linearity is seen for single-crystal paramagnetic Gd.\cite{lbdata} This behavior makes the definition of SDR ambiguous and calls for the calculation of the total resistivity in the presence of both spin and phonon scattering. This is the purpose of this section.

\begin{table}
\caption{The dataset and temperature ranges used for the fits in
Fig.\ \ref{pic-exptdata}a and the resulting slopes and
intercepts.}\label{table-exptfedata}
\begin{tabular}{|c|c|c|c|c|}
\hline
Fit & Reference & $T$ Range & Slope & Intercept \\
\# & & $(K)$ & $(\mu \Omega \text{ cm}/K)$ & $(\mu \Omega \text{ cm})$ \\
\hline 1 & Pallister & $1223 - 1523$  & $0.0304$ & $77.1$ \\
\hline 2 & Cezairliyan \emph{et.~al.} & $1500 - 1660$  & $0.0218$ & $88.0$ \\
\hline 3 & Cezairliyan \emph{et.~al.} & $1700 - 1800$  & $0.0150$ & $100$ \\
\hline
\end{tabular}
\end{table}

\begin{figure*}[htb]
\centering \includegraphics[width=0.95\textwidth]{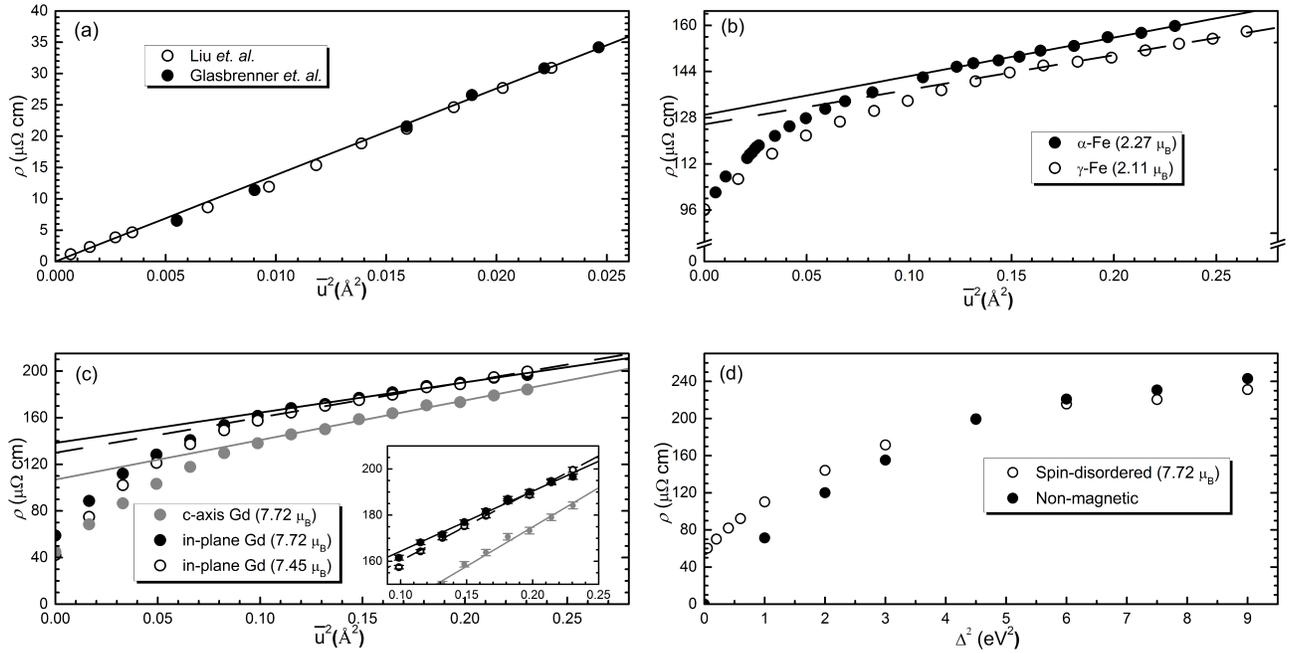} \caption{Calculated resistivities for $\alpha$ and $\gamma$ Fe and hcp Gd. (a) $\rho(\bar u^2)$ for ferromagnetic $\alpha$-Fe with atomic displacements. Filled circles: this work; open circles: data of Ref.\ \onlinecite{kellyphonons}.
(b) $\rho(\bar u^2)$ for two phases of Fe with random vector spin disorder and atomic displacements. Open circles and solid line: $\alpha$-Fe, filled circles and dashed line: $\gamma$-Fe.
(c) $\rho(\bar u^2)$ for hcp Gd with random vector spin disorder and atomic displacements. Filled gray circles and gray fit line: $c$-axis transport, $m=7.72 \mu_B$. Filled black circles and solid black fit line: in-plane transport, $m=7.72 \mu_B$. Open circles and dashed fit line: in-plane transport, $m=7.45 \mu_B$. Inset: Enlarged plot of the linear region. (d) is the resistivity as a function of square of the
Anderson disorder amplitude $\Delta^2$.
(d) a random, noncollinear, spin-disordered system is compared with a fictitious non-magnetic system.
~(d) Open circles, spin-disordered Gd ($m=7.72 \mu_B$), in-plane;
closed circles, non-magnetic Gd, in-plane.}
\label{pic-results}
\end{figure*}

Fig.\ \ref{pic-results}a shows $\rho$ as a function of $\bar u^2$ for $\alpha$-Fe calculated in the ferromagnetic state without introducing spin disorder. The linear dependence is, of course, typical since the average scattering potential is proportional to $\bar u^2$. The slope ($1381 \pm 15$ $\mu \Omega$ cm/\AA$^2$) agrees very well with the results of Liu \emph{et.\ al.} \cite{kellyphonons} obtained with a similar method.

Fig.\ \ref{pic-results}b shows the $\rho(\bar u^2)$ dependence for $\alpha$-Fe and $\gamma$-Fe with random vector spin disorder combined with atomic displacements, and Fig.~\ref{pic-results}c the results for $c$-axis and in-plane transport directions in Gd. The error bars in both panels are approximately half the height of the data symbols (0.5 and 0.9 $\mu \Omega$cm for Fe and Gd, respectively). The slopes and intercepts of the fits in Figs.~\ref{pic-results}b and \ref{pic-results}c are listed in Table~\ref{table-figuredata}.

\begin{table}
\caption{Parameters of the fits in Fig.\ \ref{pic-results}.}
\label{table-figuredata}
\begin{tabular}{|l|c|c|c|}
\hline
Element & $m$ ($\mu_B$) & Slope & Intercept \\
 &  & ($\mu \Omega$ cm/\AA$^2$) & ($\mu \Omega$ cm) \\
\hline
$\alpha$-Fe & 2.27  & $134 \pm 5$ & $129 \pm 1$ \\
\hline $\gamma$-Fe & 2.11  &  $120 \pm 6$ & $126 \pm 1$ \\
\hline Gd ($c$-axis) & 7.72  & $340 \pm 11$ & $107 \pm 2$ \\
\hline Gd (in-plane) & 7.72  & $269 \pm 9$ & $138 \pm 2$ \\
 & 7.45 & $303 \pm 9$ & $130 \pm 1$ \\
\hline
\end{tabular}
\end{table}

The values of $\bar u^2$ used in our calculations can be compared with experimental data.
Several authors extracted the temperature dependence of $\bar u^2$ from the measurements of the Debye-Waller factor for $\alpha$-Fe \cite{prakash-DW,haworth} and compared the results with models. \cite{singhDW,cavalheiro,kharoo} The experimental data for $\bar u^2$ are noisy at elevated temperatures, but the theoretical model plotted in Ref.\ \onlinecite{kharoo} may be considered as the lower bound for $\bar u^2$ at all temperatures. At the Curie temperature (1040 K) the lower bound for $\bar u^2$ is estimated at 0.053 \AA$^2$. The data for Cu \cite{pinnegarthesis} is more stable at elevated temperatures, and $\bar u^2$ at 1040 K is estimated at 0.094 \AA$^2$. For Gd the value of $\bar u^2$ at room temperature is estimated to be 0.0105 \AA$^2$, \cite{sirdeshmukhbook} while a model calculation for Er gives $\bar u^2\approx 0.169$ \AA$^2$ at its melting point.\cite{ramanand} The data used in our calculations are in line with these estimates.

The resistivity curves for $\alpha$-Fe and $\gamma$-Fe are very similar. This agrees with an experimental fact that the $\alpha$-$\gamma$ phase transition at 1180 K is barely noticeable in the resistivity plot (see Fig.\ \ref{pic-exptdata}a).

The $\rho(\bar u^2)$ curves for both Fe and Gd (Figs.\ \ref{pic-results}b and \ref{pic-results}c) deviate strongly from linearity. As $\bar u^2$ is increased, the slope decreases and eventually becomes almost constant. Below we will show that this feature is due to the breakdown of certain parts of the Fermi surface and is insensitive to the type of disorder. We interpret this as a resistivity saturation effect which takes place when the resistivity becomes of the order 100 $\mu\Omega$cm.

Clearly, Matthiessen's rule breaks down in the nonlinear regime, and the separation of the total resistivity into phonon and spin-disorder contributions becomes impossible. To facilitate further discussion, we will use the term ``bare SDR'' for the resistivity obtained at $\bar u=0$ with random spin disorder, and ``apparent SDR'' for the intercept of the linear fit to the $\rho(\bar u^2)$ curve at larger $\bar u^2$ (as listed in Table~\ref{table-figuredata}). The definition of apparent SDR is only possible as long as the slope of $\rho(\bar u^2)$ becomes approximately constant in the strong disorder regime, as it does in our calculations for Fe and Gd. The usual method of extracting the SDR from high-temperature experimental data yields the apparent SDR.

Fig.\ \ref{pic-results}b and \ref{pic-results}c show that the bare and apparent SDR are quite different in Fe and particularly in Gd. For $\alpha$-Fe ($\gamma$-Fe), the apparent SDR is 1.34 (1.32) times greater than the bare SDR, and for the $c$-axis (in-plane) transport in Gd it is 2.4 (2.3) times greater. The apparent SDR for Fe and Gd (intercepts in Table \ref{table-figuredata}) are somewhat larger than the experimental estimated of SDR obtained by extrapolating the high-temperature data (80, 108, and 96 $\mu$ $\Omega$cm for Fe, Gd with in-plane and $c$ axis transport, respectively). In the case of Fe a portion of this discrepancy is likely due to the fact that the crossover to the resistivity saturation regime is incomplete, which is strongly suggested by Fig.\ \ref{pic-exptdata}a. We also note that the inclusion of $4f$ orbitals in the basis set led to a reduction of the bare SDR \cite{alexsdr} by about 15\% and could similarly lower the apparent SDR.

For Gd the in-plane apparent SDR is 28\% greater compared to the experimental extrapolated SDR, and the $c$-axis resistivity is 11\% greater. Taking into account the experimental uncertainties, this agreement can be judged as good.

As noted earlier, \cite{jamessdr} the magnitude of the local moment has a significant effect on the bare SDR. In particular, with the local moment taken from a self-consistent CPA-DLM calculation, the bare in-plane SDR for Gd is almost 30\% lower compared to the case when the local moment is taken from the ferromagnetic calculation. We therefore performed a similar comparison for the resistivity in the presence of lattice vibrations; the corresponding curve is shown by open circles in Fig.\ \ref{pic-results}c. We observe that the difference between the resistivities calculated for $m=7.72$ and 7.45 $\mu_B$ decreases as $\bar u^2$ is increased and eventually almost disappears (see also the inset). The apparent SDR is only reduced by 6\% in the latter case, which is likely within the uncertainty of the extrapolation. This feature is consistent with the resistivity saturation phenomenon.

To gain further insight in the role of different scattering mechanisms, we repeated the calculations of the in-plane resistivity of Gd with a fictitious Anderson disorder introduced in lieu of the random lattice displacements. For comparison we considered the spin-disordered system with $m=7.72 \mu_B$ and its non-magnetic counterpart with unpolarized $4f$ cores. Anderson disorder is characterized by an amplitude $\Delta$ (see Sec.\ \ref{section-methods}), and the results are plotted in Fig.~\ref{pic-results}d as a function of $\Delta^2$.

The shape of the $\rho(\Delta^2)$ curve in Fig.\ \ref{pic-results}d for a system with spin disorder (open circles) is similar to $\rho(\bar u^2)$ for phonon disorder in Fig.\ \ref{pic-results}c. A similar curve is obtained for a non-magnetic system (filled circles Fig.\ \ref{pic-results}d) with an obvious exception that the curve starts from zero rather than from the bare SDR. The similarity of the resistivity curves for different types of disorder indicates that the resistivity saturation effect is primarily controlled by the features of the electronic structure. These features will be studied in the following section. Similar to the case of the phonon disorder discussed above, the two curves for spin-disordered and magnetic systems shown in Fig.\ \ref{pic-results}d approach each other at large $\Delta^2$.

We return to the high-temperature resistivity measurements \cite{vedernikov} taken on polycrystalline samples compiled in Fig.\ \ref{pic-exptdata}b and compare with our results. The shape of the curves is remarkably similar to those in Fig.\ \ref{pic-results}d. First, the resistivity saturation trend is clearly seen for all elements including the nonmagnetic lutecium, with deviations from linearity setting in when the resistivity exceeds about 100 $\mu\Omega$cm. Second, while the intercept of the resistivity steadily increases with the magnitude of the spin magnetic moment (i.\ e.\ with the decreasing atomic number), the curves tend to converge at high temperatures. Our results are in excellent agreement with both of these features. By comparing the total resistivities, we can also estimate that at $T\sim1000$ K the magnitude of lattice disorder $\bar u\sim 0.4$ \AA, and a similar relaxation rate is generated by Anderson disorder with $\Delta\sim1.8$ eV.

\section{Disorder-induced partial Fermi surface collapse in G\lowercase{d}}
\label{section-electronicstructure}

\begin{figure}[htb]
\centering \includegraphics[width=0.45\textwidth]{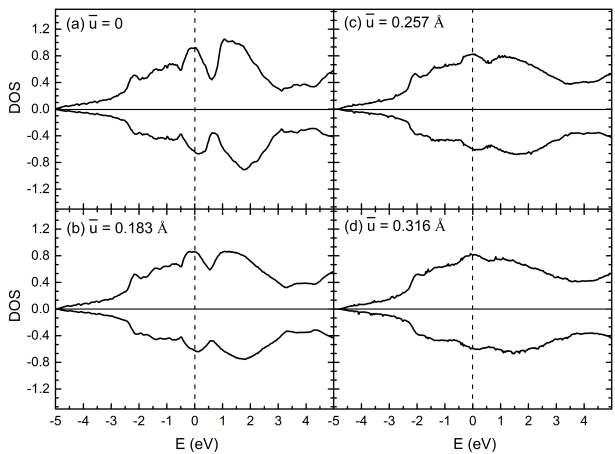} \caption{Average spin-projected
local density of states of spin-disordered Gd ($m=7.72 \mu_B$) for
different amplitudes of lattice disorder $\bar u$: (a) No phonon disorder, (b) $\bar u=0.183$ \AA, (c) $\bar u=0.257$\AA,
(d) $\bar u=0.316$ \AA.} \label{pic-dos}
\end{figure}

In order to understand the origin of the resistivity saturation effect, in this section we analyze the influence of disorder on the electronic structure of Gd. First, let us examine the evolution of the DOS in spin-disordered Gd as the lattice disorder is increased, which is presented in Fig.\ \ref{pic-dos} (see Sec.\ \ref{section-methods}). At $\bar u$ the DOS is the same as in Ref.\ \onlinecite{jamessdr} and similar to the DLM calculation.\cite{khmelevskyiDLM2} Although spin disorder smears out the sharp variations of the DOS, one can still see pronounced features. As lattice disorder is introduced, these features are also gradually smeared out. The suppression of the DOS structure correlates with the reduction of the slope of the resistivity in Fig.~\ref{pic-results}c.

\begin{figure}[htb]
\begin{center}
\includegraphics[width=0.45\textwidth]{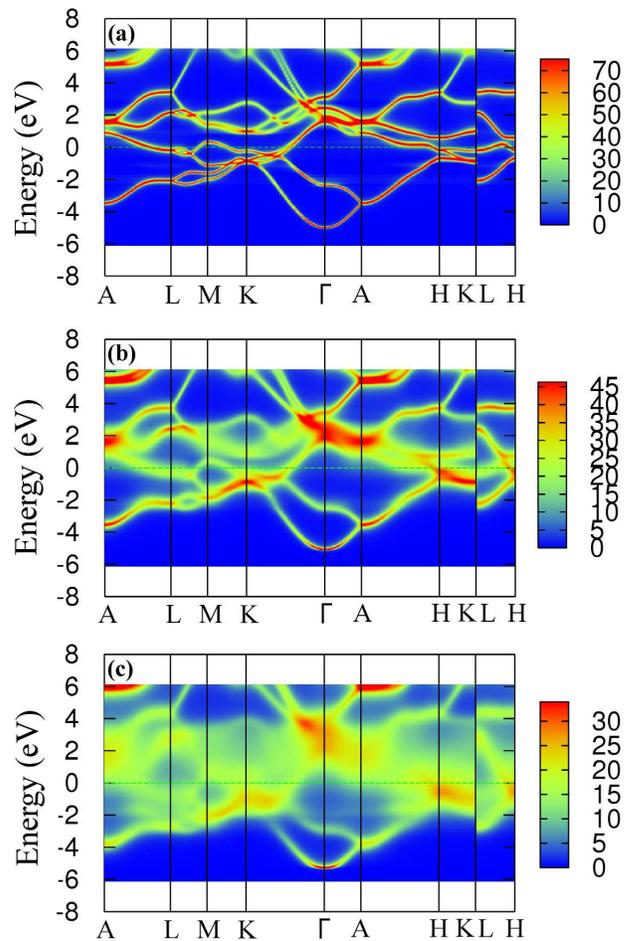}
\end{center}
\caption{The
spectral function of paramagnetic Gd for different amplitudes $\Delta$ of Anderson
disorder plotted along high-symmetry lines of the hexagonal Brillouin zone. (a)
$\Delta=0$. (b) $\Delta=0.95$ eV. (c) $\Delta = 1.8$  eV.}
\label{pic-symspectral}
\end{figure}

\begin{figure}[htb]
\centering \includegraphics[width=0.45\textwidth]{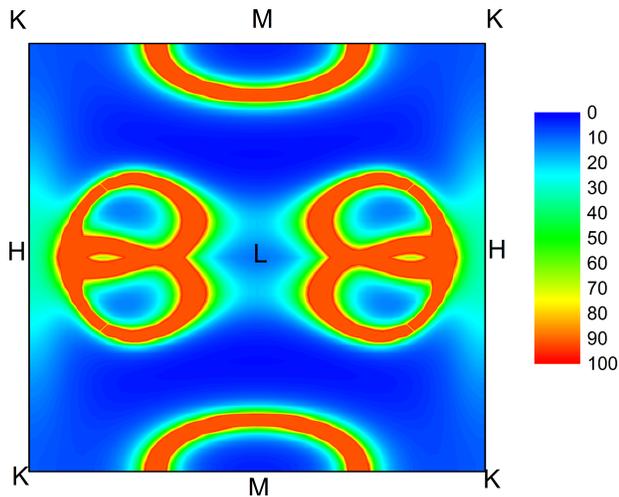} \caption{The spectral
function of paramagnetic Gd evaluated at the Fermi energy for the HLMK plane in
the Brillouin zone.} \label{pic-hlmk}
\end{figure}

\begin{figure*}[htb]
\centering \includegraphics[width=0.95\textwidth]{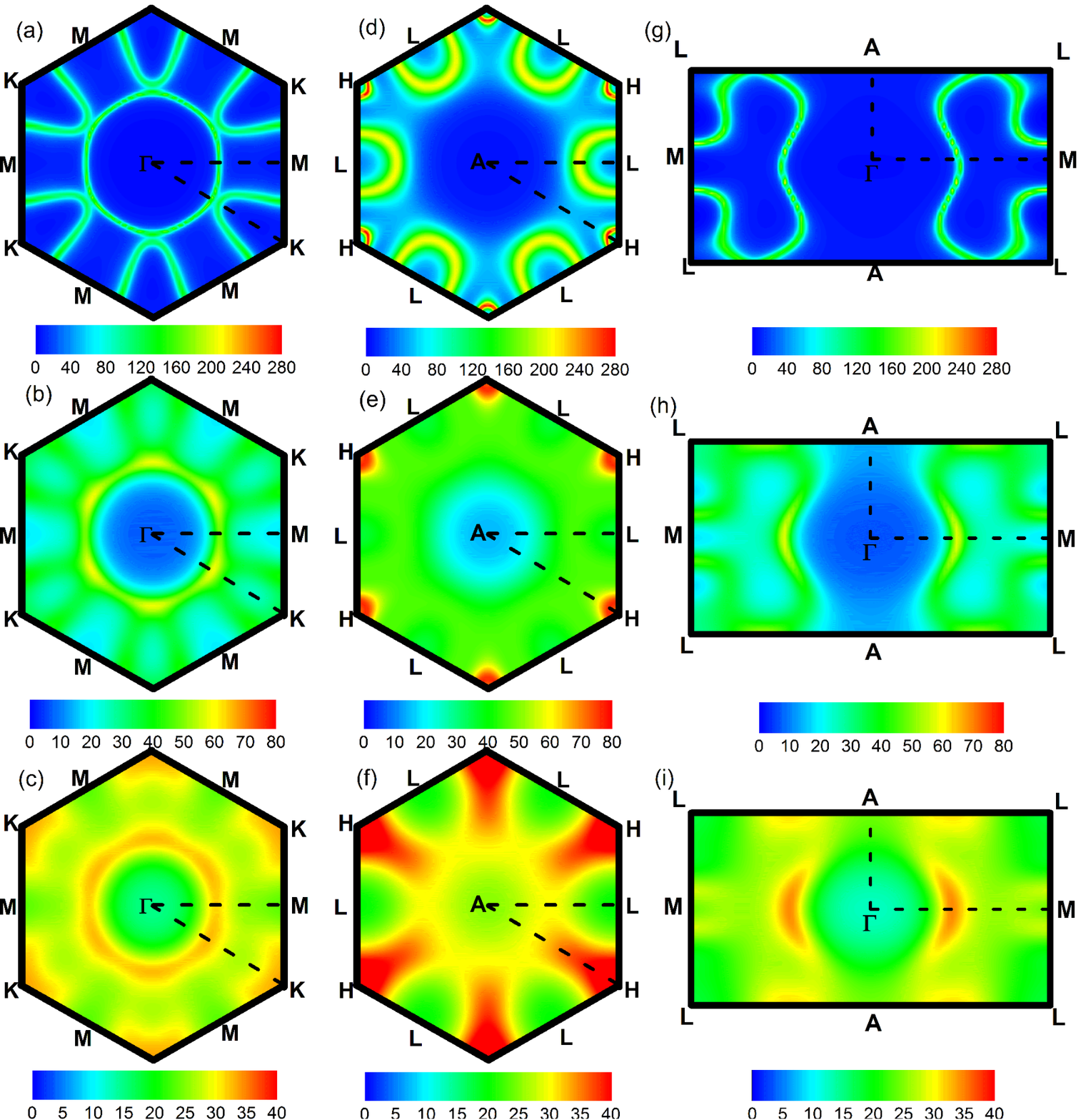}
\caption{The spectral function of paramagnetic Gd evaluated at the Fermi energy on the indicated planes of the Brillouin zone for different values of the Anderson disorder amplitude $\Delta$. (a), (d), (g) $\Delta=0$. (b), (e), (h) $\Delta=0.95$ eV. (c), (f), (i) $\Delta=1.8$ eV.} \label{phsdr-pic-spectral}
\end{figure*}

Further, we have calculated the Bloch spectral function for paramagnetic Gd including Anderson disorder of a varying amplitude (see Section \ref{section-methods}). Anderson disorder is used instead of lattice disorder in order to simplify its treatment within CPA.
Fig.~\ref{pic-symspectral} shows the energy-dependent spectral function plotted along several high-symmetry lines in the Brillouin zone. The three panels represent different disorder amplitudes. In addition, Figs.\ \ref{pic-hlmk} and \ref{phsdr-pic-spectral} display several slices of the spectral function at the Fermi energy.

The spectral function of paramagnetic Gd without lattice disorder (Fig.~\ref{pic-symspectral}a) shows that it has a well-defined, weakly broadened Fermi surface, and that the exchange splitting is completely absent. This corresponds to the Stoner picture, which is consistent with several photoemission experiments \cite{dowbengd,erskineharmongd,passekgd} and
calculations,\cite{noltinggd}
although this conclusion has been controversial.\cite{maiti} Note that although the band structure is very similar to that
of a fictional material with unpolarized $4f$ states, it coexists with fluctuating local magnetic moments and with the exchange splitting of the local DOS shown in Fig.\ \ref{pic-dos}a.

The Fermi surface is also readily identified in Fig.~\ref{pic-hlmk} and panels (a), (d), and (g) of Fig.\ \ref{phsdr-pic-spectral} which correspond to pure spin disorder. This Fermi surface has a hole-like cylindrical sheet centered around the $\Gamma-\text{A}$ line and an electron-like sheet outside it.\cite{keetonloucks,coqblin1977} There are
several points where the electron and hole sheets approach each other, such as along the $\Gamma-\text{K}$ line; the sheets cross near the $\Gamma-\text{H}$ line and are degenerate everywhere on the AHL plane.

The spectral function in Fig.\ \ref{pic-hlmk} is plotted in the same cross-section as Fig.\ 2 in Ref.\ \onlinecite{hughesnaturelett}, which was obtained using the self-interaction-corrected LSDA. Although the Fermi surface features appearing in these plots are immediately identifiable with each other, there are notable differences in their shapes. These differences are due to the different approximations used in the description of the $4f$ electrons. They are immaterial to the general conclusions that follow.

As the Anderson disorder amplitude is increased (Fig.\ \ref{pic-symspectral}a-c), the bands broaden, and eventually a large portion of the Fermi surface is destroyed. This evolution can also be clearly observed in Fig.\ \ref{phsdr-pic-spectral} showing the Brillouin zone cuts at the Fermi energy. The second row of panels (b, e, and h) in Fig.\ \ref{phsdr-pic-spectral} corresponds to the same disorder amplitude as Fig.\ \ref{pic-symspectral}b, and the third row (c, f, and i) to the same amplitud as Fig.\ \ref{pic-symspectral}c. For $\Delta = 0.95 \text{ eV}$, disorder has a much stronger effect on the Fermi surface close to the ALH plane compared to the remainder of the Brillouin zone. Only a portion of the hole-like Fermi surface sheet near the $\Gamma$MK plane survives in the presence of disorder. The states near the ALH plane are strongly affected due to the degeneracy of the electron-like and hole-like sheets on this plane, which are therefore strongly mixed by disorder. In addition, the surviving part of the Fermi surface corresponds to the bands with a higher Fermi velocity (see Fig.\ \ref{phsdr-pic-spectral}), which reduces the extent of the broadening in $k$-space observed at a given energy. For $\Delta = 1.8 \text{ eV}$, the few remaining features of the Fermi surface are destroyed and an incoherent spectral weight spans the entire Brillouin zone.

The collapse of a large portion of the Fermi surface correlates with the large decrease in the slope of the resistivity in Fig.\ \ref{pic-results}d, giving additional support to the interpretation of these results as a resistivity saturation effect.

The results in Fig.\ \ref{phsdr-pic-spectral} can also help reconcile the recent angle-resolved photoemission spectroscopy (ARPES) measurements for paramagnetic Gd \cite{prb-81-012401} with the calculated Fermi surface of non-magnetic (or spin-disordered) Gd.\cite{keetonloucks,coqblin1977} At room temperature ARPES only reveals a corrugated-cylinder feature, while the theoretical Fermi surface also has complicated features centered at the AHL plane of the Brillouin zone. As discussed above, disorder strongly broadens the spectral features near this plane due to the presence of degeneracy. This suggests that lattice vibrations may suppress the additional features of the Fermi surface and make them indiscernible in ARPES. The ARPES signal in the $\Gamma$MLA plane has a diffuse ``halo'' outside of the cylindrical sheet, and its shape is in reasonable agreement with Fig.\ \ref{phsdr-pic-spectral}h. Thus, the presence of a diffuse scattering region instead of a sharp electron-like sheet in ARPES measurements may be due to disorder-induced band broadening.

Although we have only calculated the resistivity of Gd in the presence of both spin and lattice disorder, we can consider the implications of the results for the whole rare-earth series. The issue of quantum corrections is of particular interest. If the $4f$ orbitals are treated as fully localized with a well-defined total angular momentum $J$ (strong spin-orbit coupling limit), the resistivity in the paramagnetic state should be proportional \cite{deGennes,Brout,Kasuya} to the so-called de Gennes factor $(g-1)^2J(J+1)$.\cite{Jensen} This factor takes into account the quantum structure of the $J$ mutliplet. The analysis of early experimental data \cite{Legvold} suggested that the out-of-plane resistivity in the Gd-Tm series scales with the de Gennes factor, while the in-plane resistivity scales with $S(S+1)$. In order to reconcile this unexpected trend with the model, Legvold \cite{Legvold} claimed that the $S(S+1)$ scaling is accidental and introduced an empirical correction based on the slope $d\rho/dT$ of the resistivity above the magnetic transition temperature, assuming that the large factor-of-two variation of this slope through the Gd-Tm series is due to the changes in the Fermi surface area. However, the calculated variation in the relevant Fermi-velocity integral across the series is only about 20\%, \cite{jamessdr} which is too small compared with the observed variation of $d\rho/dT$. The results presented above along with the high-temperature resistivity measurements \cite{vedernikov} show that the variation in $d\rho/dT$ is largely due to the resistivity saturation trend and not to the changes in the Fermi surface.

As regards the absolute values of the resistivity, we found that the comparison with experimental data for lighter elements in the Gd-Tm series requires that lattice disorder is included in the calculation along with spin disorder. At least for Gd the resistivity calculated in this way is in reasonable agreement with experiment. For heavier elements with lower transition temperatures, saturation effects remain insignificant in the region used for the fitting, and the SDR extracted from experiment can therefore be directly compared with the calculated bare SDR. For these heavier elements the agreement with experiment appears to be significantly improved by assuming $S(S+1)$ scaling.\cite{jamessdr} This kind of scaling occurs if the spin and orbital moments are not strongly coupled to each other, which is surprising for heavy rare-earth elements. While understanding the origin of this behavior is beyond the scope of this paper, we suggest that the finite width of the $4f$ band, if comparable to the spin-orbit multiplet splitting, can destroy the strong correlation between the spin and orbital moment. This issue requires further investigation.

\section{Conclusions}
\label{section-conclusions}

We have analyzed the resistivity of $\alpha$-Fe, $\gamma$-Fe, and hcp Gd in the presence of both spin and lattice disorder.
Strong deviations from Matthiessen rule were found. As the resistivity approaches values of order 100 $\mu\Omega$ cm, resistivity saturation effects start to manifest themselves. When plotted against the square of the disorder amplitude, the resistivity crosses over into a high-disorder regime with a much smaller slope, which tends to approach a constant. These results are in excellent agreement with high-temperature resistivity data for rare-earth metals.

Extrapolation from the quasi-linear region in the paramagnetic state leads to an ``apparent'' spin-disorder resistivity (SDR) which exceeds the ``bare'' SDR (calculated without lattice disorder) by a factor 2.4 in Gd and 1.3 in both phases of Fe. Thus, taking lattice disorder into account resolves the large discrepancy between earlier calculations of SDR with experimental data for Gd. By analyzing the spectral functions in the presence of disorder, we have argued that the resistivity saturation in Gd is due to the collapse of a large portion of the Fermi surface, which is promoted by the degeneracy of the electron and hole-like sheets at the ALH plane in the Brillouin zone.

\acknowledgments

This research was supported by the National Science Foundation through Grant No. DMR-1005642, the Nebraska MRSEC (DMR-0820521), and in part under Grant No. NSF PHY11-25915. Computations were performed utilizing the Holland Computing Center at the University of Nebraska. 


%

\end{document}